\documentclass[12pt,a4paper]{iopart}

\usepackage{iopams}
\usepackage{setstack}
\usepackage{aas_macros}

\usepackage{url}
\usepackage{graphicx}
\usepackage{siunitx}
\usepackage{booktabs}

\usepackage[numbers]{natbib}

\begin{document}

\title{Modelling the response of potassium vapour in resonance scattering spectroscopy}

\author{S J Hale$^{1,2}$, W J Chaplin$^{1,2}$, G R Davies$^{1,2}$, Y P Elsworth$^{1,2}$}
\address{$^1$ School of Physics and Astronomy, University of
  Birmingham, Edgbaston, Birmingham B15 2TT, United Kingdom}
\address{$^2$ Stellar Astrophysics Centre, Department of Physics and
  Astronomy, Aarhus University, Ny Munkegade 120, DK-8000 Aarhus C,
  Denmark}
\ead{s.j.hale@bham.ac.uk}
\begin{indented}

\item[]{}\quad\rm\ignorespaces

This is an author-created, un-copyedited version of an article
accepted for publication in Journal of Physics B. The publisher is not
responsible for any errors or omissions in this version of the
manuscript or any version derived from it.
\end{indented}

\begin{abstract}
Resonance scattering techniques are often used to study the properties
of atoms and molecules.  The Birmingham Solar Oscillations Network
(BiSON) makes use of Resonance Scattering Spectroscopy by applying the
known properties of potassium vapour to achieve ultra-precise Doppler
velocity observations of oscillations of the Sun.  We present a model
of the resonance scattering properties of potassium vapour which can
be used to determine the ideal operating vapour temperature and
detector parameters within a spectrophotometer.  The model is
validated against a typical BiSON vapour cell using a tunable diode
laser, where the model is fitted to observed absorption profiles at a
range of temperatures.  Finally we demonstrate using the model to
determine the effects of varying scattering detector aperture size,
and vapour temperature, and again validate against observed scattering
profiles.  Such information is essential when designing the next
generation of BiSON spectrophotometers ({BiSON:NG}), where the aim is
to make use of off-the-shelf components to simplify and miniaturise
the instrumentation as much as practical.
\end{abstract}


\maketitle

\ioptwocol

\section{Introduction}

Resonance scattering techniques are often used to study the properties
of atoms and molecules, such as energy levels and the fine and
hyperfine structure~\citep{doi:10.1098/rsta.1979.0091}.  When applied
to absorption spectroscopy with techniques such as optical pumping
(so-called Saturated Absorption Spectroscopy) it is possible to
achieve a resolution that is Doppler-free and limited only by the
uncertainty principle, producing a highly stable locking reference for
tunable laser systems~\citep{doi:10.1119/1.18457,Svanberg2004}.
Resonance scattering spectroscopy has wide reaching multi-disciplinary
applications including biomedical diagnostics~\citep{Svanberg:16}, and
environmental sensing in terms of atmospheric
pollution~\citep{Svanberg:16}.  In astronomy, vapour reference cells
are often used to enable the high-precision spectrograph wavelength
calibration required for asteroseismology and detection of
exoplanets~\citep{Grundahl_2017, 10.1117/12.2056668}.

The Birmingham Solar Oscillations Network (BiSON) makes use of
resonance scattering with a potassium vapour reference cell, enabling
ultra-precise Doppler velocity measurements of oscillations of the
Sun~\citep{s11207-015-0810-0}.  In this article we present a model of
the resonance scattering properties of potassium vapour which can be
used to determine the ideal operating vapour temperature, and
spectrophotometer detector parameters, which are essential when
developing new instrumentation making use of off-the-shelf
components~\citep{halephd}.

In Section~\ref{s:instrumentation} we give a brief overview of BiSON
instrumentation and the use of resonance scattering spectroscopy.  In
Section~\ref{s:resonance} we review the theory of spectral line
formation from first principles, beginning from a similar position to
\citet{1993ExA.....4...87B} for sodium vapour, and in
Section~\ref{s:opticaldepth} develop the equations necessary to
determine optical depth from vapour temperature.  In
Sections~\ref{s:calibration} and~\ref{s:fitting} we use a tunable
diode laser to validate the model against the absorption profiles
observed with a standard BiSON vapour cell at a range of temperatures.
Finally in Section~\ref{s:detector} we use the model to determine the
ideal vapour temperature and detector configuration to maximise the
detected scattering intensity, and again validate against observed
scattering profiles.

\section{BiSON Instrumentation}
\label{s:instrumentation}

\begin{figure*}[t]
  \centering
  \includegraphics[scale=1.0]{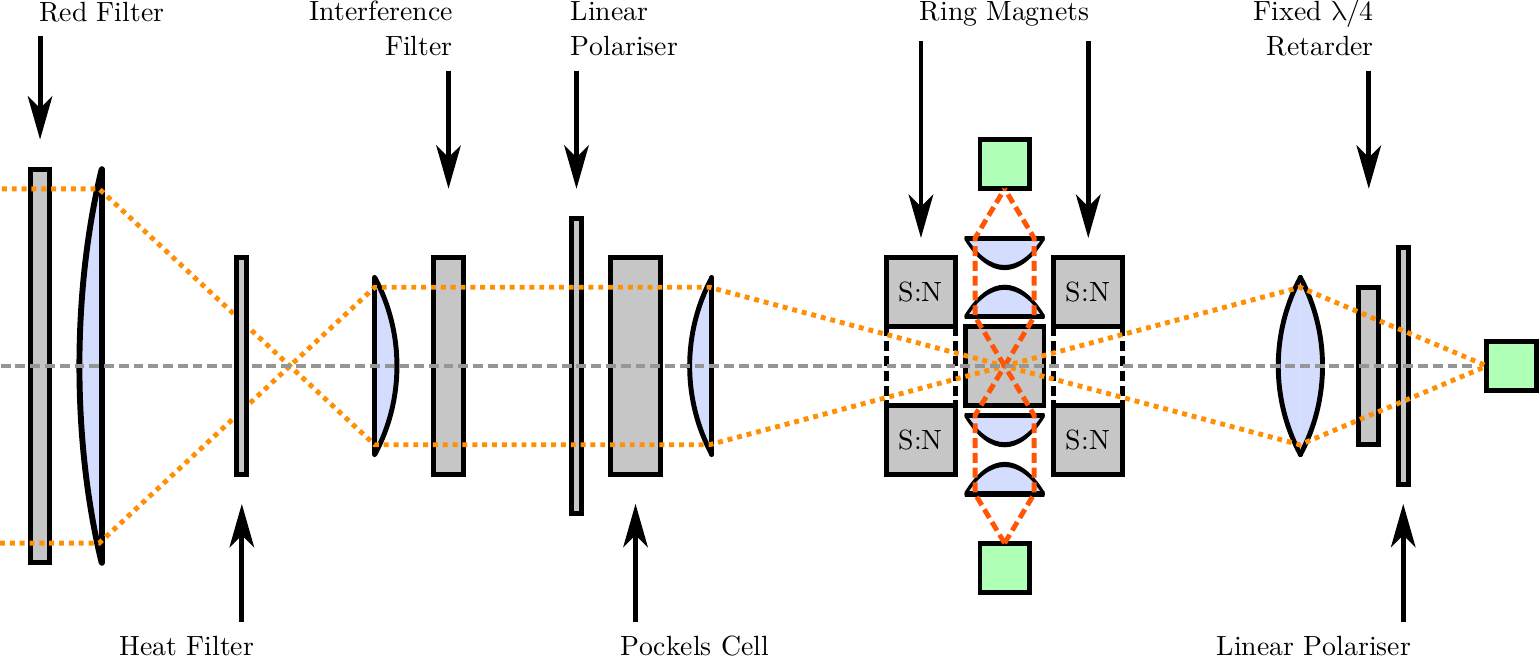}
  \caption{Schematic of a typical modern BiSON resonance scattering
    spectrophotometer.  Light travels from left to right.  The green
    boxes indicate detector locations.  The magnetic field across the
    vapour cell is longitudinal, with the field axis aligned with the
    optical axis.}
  \label{fig:bison_rss}
\end{figure*}

The original BiSON resonance scattering spectrophotometer (RSS)
developed for Doppler velocity astronomical observations is described
by~\citet{1978MNRAS.185....1B}.  Modern BiSON RSS follow the same
principles but make use of slightly different designs, and a typical
example is shown in Figure~\ref{fig:bison_rss}.  Light from the Sun is
first passed through a~\SI{30}{\milli\metre} diameter aperture, and a
pair of lenses in a Keplerian telescope arrangement are used to
compress the beam reducing the diameter in preparation for subsequent
smaller optics.  A red long-pass filter, and an infra-red short-pass
filter, form a pair to relieve the thermal load on the
instrumentation.  An interference filter with a bandwidth
of~\SI{1.5}{\nano\metre} is used to isolate only the potassium D1
Fraunhofer line at~\SI{769.898}{\nano\metre}, excluding the D2
line. The potassium is contained within a small cubic glass cell with
sides of length~\SI{17}{\milli\metre}, and a short stem.  The stem is
typically heated to around~\SI{90}{\degreeCelsius} to cause the
formation of potassium vapour within the cell, with the cubic cell
itself maintained at a temperature higher than this to ensure the
glass remains clear of solid potassium.  The cell is placed in a
longitudinal magnetic field which causes the absorption line to be
Zeeman split into two components, where their separation is dependent
on the magnetic field strength.  Splitting the lab reference-frame in
this way allows measurements to be taken at two working points on the
corresponding absorption line in the solar atmosphere.  The
polarisation component is selected using a linear polariser and an
electro-optic Pockels-effect quarter-wave retarder.  The polarisation
is switched at approximately~\SI{100}{\hertz} and a normalised ratio,
R, formed,
\begin{equation}
  R = \frac{I_{\mathrm{b}} - I_{\mathrm{r}}}{I_{\mathrm{b}} + I_{\mathrm{r}}} ~,
\end{equation}
where $I_{\mathrm{b}}$ and $I_{\mathrm{r}}$ are the measured
intensities in the blue and red wings respectively.  Normalising the
ratio in this way reduces the effect of atmospheric scintillation
allowing low-noise photometry through Earth's
atmosphere~\citep{2020PASP..132c4501H}, and also means it is
unnecessary to know the precise value of the magnetic field strength
or the precise frequency of the potassium line on the Sun.  Light is
focused into the centre of the potassium cell, where it is resonantly
scattered by the vapour.  Light that interacts with the vapour is
scattered isotropically, and it is this isotropic nature that is
leveraged by the {BiSON} resonance scattering spectrophotometers to
move a photon away from the instrumental optical axis and into a
detector. Two photodiode detectors on either side of the beam return a
precise measurement of the changing solar radial velocity.  A third
detector monitors unscattered light that passes directly through the
instrument.

There are several natural broadening processes which affect the
properties of an absorption line, and it is necessary to understand
the environmental factors (temperature and pressure) that affect these
processes when building a model of expected instrumentation
performance and informing new designs~\citep{halephd}.

\section{Resonance Radiation}
\label{s:resonance}

\subsection{Natural Line Width}

In a gas that forms an absorption line, the absorption coefficient
$\kappa_\nu$ of the gas is defined in terms of,
\begin{equation}\label{eq:absorptioncoeff}
  I_\nu = I_0 e^{-\kappa_\nu x} ~,
\end{equation}
where $I_0$ is the intensity of incident light, $I_\nu$ is the
intensity of the transmitted light, and $x$ is the thickness of the
absorbing material.  The units of the absorption coefficient are
always expressed in terms of the reciprocal of the units of $x$ such
that the exponent becomes a unitless ratio.

It can be shown~\citep[p.~95]{mitchell1961resonance} that the integral
of $\kappa_\nu$ over all frequencies is equal to,
\begin{equation}\label{eq:intnatural1}
  \int \kappa_\nu \mathrm{d} \nu =
    \frac{\lambda_0^2}{8\pi}
    \frac{g_2}{g_1}
    \frac{N}{\tau}
    \left( 1 - \frac{g_1}{g_2} \frac{N'}{N} \right) ~,
\end{equation}
where $\lambda_0$ is the wavelength at the centre of the absorption
line, $\tau$ the lifetime of the excited state, and $g_1$ and $g_2$
are the statistical weights of the normal and excited states
respectively.  $N$ is the number density of atoms of which
$[N-N']$ are capable of absorbing in the frequency range $\nu$ to
$\nu~+~\mathrm{d}\nu$, and similarly $N'$ is the number of excited
atoms that are capable of emitting in this range.  In the case where
the ratio of excited atoms to normal atoms is very small
($\leqslant~10^{-4}$) then $N'$ can be neglected and
equation~\ref{eq:intnatural1} reduces to,
\begin{equation}\label{eq:intnatural2}
  \int \kappa_\nu \mathrm{d} \nu =
    \frac{\lambda_0^2}{8\pi}
    \frac{g_2}{g_1}
    \frac{N}{\tau} ~,
\end{equation}
and this result shows that the integral of the absorption coefficient
is constant if $N$ remains constant.  This assumption can normally be
made when the formation of excited atoms is due to absorption of a
beam of light.  Where a gas is electrically excited at high current
densities, for example in a gas discharge lamp, the number of excited
atoms may become a high fraction of the total number and so the
simplification in equation~\ref{eq:intnatural2} cannot be made.  In
the case where all atoms capable of absorbing at a certain frequency
are already excited, the gas is said to be saturated and no further
absorption can take place at that frequency.

Although the energy of an atomic transition is fixed, and therefore
similarly the frequency of light with which it can interact, the
absorption profile has a natural line width which arises from the
uncertainty principle.  When considered in the form,
\begin{equation}
  \Delta E \Delta t > \frac{\hbar}{2} ~,
\end{equation}
where $\Delta E$ is the energy, $\Delta t$ the lifetime, and $\hbar$
the reduced Planck constant, the result suggests that for extremely
short excited state lifetimes there will be a significant uncertainty
in the energy of the photon emitted.  If the energy of many photons
are measured the distribution formed is Lorentzian in shape, where
$\Gamma$ is the width parameter for a Lorentzian profile. If we assume
the lifetime in the excited state is the uncertainty in time (i.e.,
$\tau = \Delta t$) then,
\begin{equation}\label{eq:up2}
  \Delta E = \frac{\Gamma}{2} = \frac{\hbar}{2 \tau} ~,
\end{equation}
and so,
\begin{equation}\label{eq:up3}
  \Gamma = \frac{\hbar}{\tau} ~,
\end{equation}
which shows that the natural line width is inversely
proportional to the lifetime of the atom in the excited
state.  \citet{PhysRev.77.153} observed the natural lifetime of the
potassium {$4^2P_{\frac{1}{2}}$\,--\,$4^2S_{\frac{1}{2}}$} transition
to be~\SI{27}{\nano\second}.  Using equation~\ref{eq:up3} this
corresponds to a Lorentzian width of~\SI{2.44e-8}{\electronvolt}.  We
can convert this to frequency using the relation $E = h\nu$, and so
equation~\ref{eq:up2} becomes,
\begin{equation}
  \Delta \nu   = \frac{1}{4 \pi \tau} ~,
\end{equation}
producing a width of~\SI{5.9}{\mega\hertz}.  For the line centre of
the {$4^2P_{\frac{1}{2}}$\,--\,$4^2S_{\frac{1}{2}}$} atomic transition
at~\SI{769.898}{\nano\meter} this is a full width at half maximum
({FWHM}) of just~\SI{1.166e-05}{\nano\meter}, which is equivalent to a
Doppler velocity of~\SI{4.54}{\meter\per\second}.


\subsection{Pressure Broadening}

Pressure acts to modify the natural line width.  When an excited atom
collides with another particle, it can be stimulated to decay earlier
than would be the case for its natural lifetime.  Continuing the
argument from the uncertainty principle as before, if the lifetime is
reduced than the uncertainty on the energy emitted must increase and
hence broaden the line width.  As the pressure increases, the
likelihood of collisions increases.

Experimental conditions are usually selected to ensure that pressure
broadening is minimised to a level of insignificance, and this is the
case for our vapour reference cells which are placed under vacuum
before being filled with their reference element.  We will neglect
this effect.


\subsection{Doppler Broadening}

Since the atoms in a gas are moving, the wavelength of light absorbed
or emitted by the gas will be shifted according to the standard
non-relativistic Doppler formula,
\begin{equation}
  \frac{\Delta \lambda}{\lambda_0} = -\frac{\Delta \nu}{\nu_0} = \frac{v}{c} ~,
\end{equation}
where $\Delta\lambda$ is the shifted wavelength, $\lambda_0$ is the
rest wavelength, $v$ the speed, and $c$ the speed of light.
Similarly, $\Delta \nu$ is the shift in frequency, and $\nu_0$ is the
frequency at rest.  The atoms in a gas are moving at speeds defined by
the Maxwell-Boltzmann distribution, and so a spectral line will be
broadened into a range of possible wavelengths with a Gaussian
distribution. The FWHM due to Doppler broadening is given by,
\begin{equation}\label{eq:doppler6}
  \Delta \nu = \sqrt{\frac{8 k_\mathrm{B} T \ln{2}}{mc^2}} \nu_0
\end{equation}
which can be used to calculate the expected Doppler width of the
absorption line.  Potassium has a standard atomic weight
of~\SI{39.0983}{\atomicmassunit}.  For a reference vapour
at~\SI{100}{\celsius} the broadening is~\SI{0.0017}{\nano\meter}.
This is equivalent to a Doppler velocity
of~\SI{663}{\meter\per\second}, almost 150~times larger than the width
produced by natural broadening.  On the Sun at a temperature
of~\SI{5777}{\kelvin}, the D1 line width due to Doppler broadening
is~\SI{0.0067}{\nano\meter}.

It may be shown~\citep[p.~99]{mitchell1961resonance} that when
considering only Doppler broadening, the absorption coefficient of a
gas is given by,
\begin{equation}\label{eq:dopplerabs}
  \kappa_\nu = \kappa_0 e^{- \left[ \frac{2(\nu - \nu_0)}{\Delta \nu} \sqrt{\ln{2}} \right]^2} ~,
\end{equation}
where $\kappa_0$ is the ideal maximum absorption for Doppler
broadening alone, and $\Delta \nu$ is the Doppler breadth defined in
equation~\ref{eq:doppler6}.  The integral of $\kappa_\nu$ over all
frequencies is equal to,
\begin{equation}\label{eq:intdoppler}
  \int \kappa_\nu \mathrm{d} \nu
    = \frac{1}{2} \sqrt{\frac{\pi}{\ln{2}}} \kappa_0 \Delta \nu ~,
\end{equation}
and this depends only on the temperature and the atomic mass of the
atoms forming the vapour, which for a given material means temperature
is the only variable.

\section{Vapour Optical Depth}
\label{s:opticaldepth}

To be able to interpret the results of any experiments involving a gas
reference cell, we need a mathematical expression for the absorption
of the gas under the combined broadening conditions.  We know that no
matter what physical processes are responsible for the formation of a
spectral line, the integral of the absorption coefficient over all
frequencies must remain constant if the number of atoms remains
constant.  This can be explained more simply by considering that any
given atom can have the appropriate physical properties (i.e., speed,
direction) to interact with only one particular frequency.  If a
spectral line is broadened, then the absorption depth must
simultaneously reduce due to there being fewer atoms available to
interact at the line centre.  From this we can equate
equations~\ref{eq:intnatural2} and~\ref{eq:intdoppler} thus,
\begin{equation}
    \frac{\lambda_0^2}{8\pi}
    \frac{g_2}{g_1}
    \frac{N}{\tau}
    =
    \frac{1}{2}
    \sqrt{\frac{\pi}{\ln{2}}}
    \kappa_0 \Delta \nu ~,
\end{equation}
and solving for $\kappa_0$ produces~\cite[p.~100]{mitchell1961resonance},
\begin{equation}\label{eq:k0}
  \kappa_0 = \frac{2}{\Delta \nu}
             \sqrt{\frac{\ln{2}}{\pi}}
             \frac{\lambda^2_0}{8 \pi}
             \frac{g_2}{g_1}
             \frac{N}{\tau} ~,
\end{equation}
which is the maximum absorption possible.  This can be combined with
the expected line profile to produce the final value for $\kappa_\nu$.
We have already seen such an example in equation~\ref{eq:dopplerabs}
where for a line profile dominated by Doppler broadening $\kappa_\nu$
has a Gaussian profile.

The absorption coefficient and thickness in
equation~\ref{eq:absorptioncoeff} can be combined into one term
$\tau_\nu$, known as the optical depth,
\begin{equation}\label{eq:od}
  I_\nu = I_0 e^{-\tau_{\nu}} ~,
\end{equation}
where, as before, $I_0$ is the intensity of incident light, and
$I_\nu$ is the intensity of the transmitted light.  Unit optical depth
is therefore defined as the optical thickness that reduces the photon
flux by a factor of $1/e$.  It should be noted that $\tau_\nu$ is
specifically a different quantity from $\tau$ the lifetime of the
excited state.

We know that the \SI{769.898}{\nano\meter} line profile formed by
potassium vapour is dominated by Doppler broadening, and so
equation~\ref{eq:dopplerabs} is valid to describe the distribution.
In order to evaluate $\kappa_\nu$ we need to know the value of all the
terms in equation~\ref{eq:k0} -- temperature $T$, the atomic mass $m$,
the frequency and wavelength of the line centre $\nu_0$ and
$\lambda_0$, the number density of absorbing atoms $N$, the lifetime
of the excited state $\tau$, and the degeneracy of the
{$4^2P_{\frac{1}{2}}$\,--\,$4^2S_{\frac{1}{2}}$} energy levels $g_1$
and $g_2$, respectively.

\begin{figure*}[t]%
  \centering
  \includegraphics[scale=1.0]{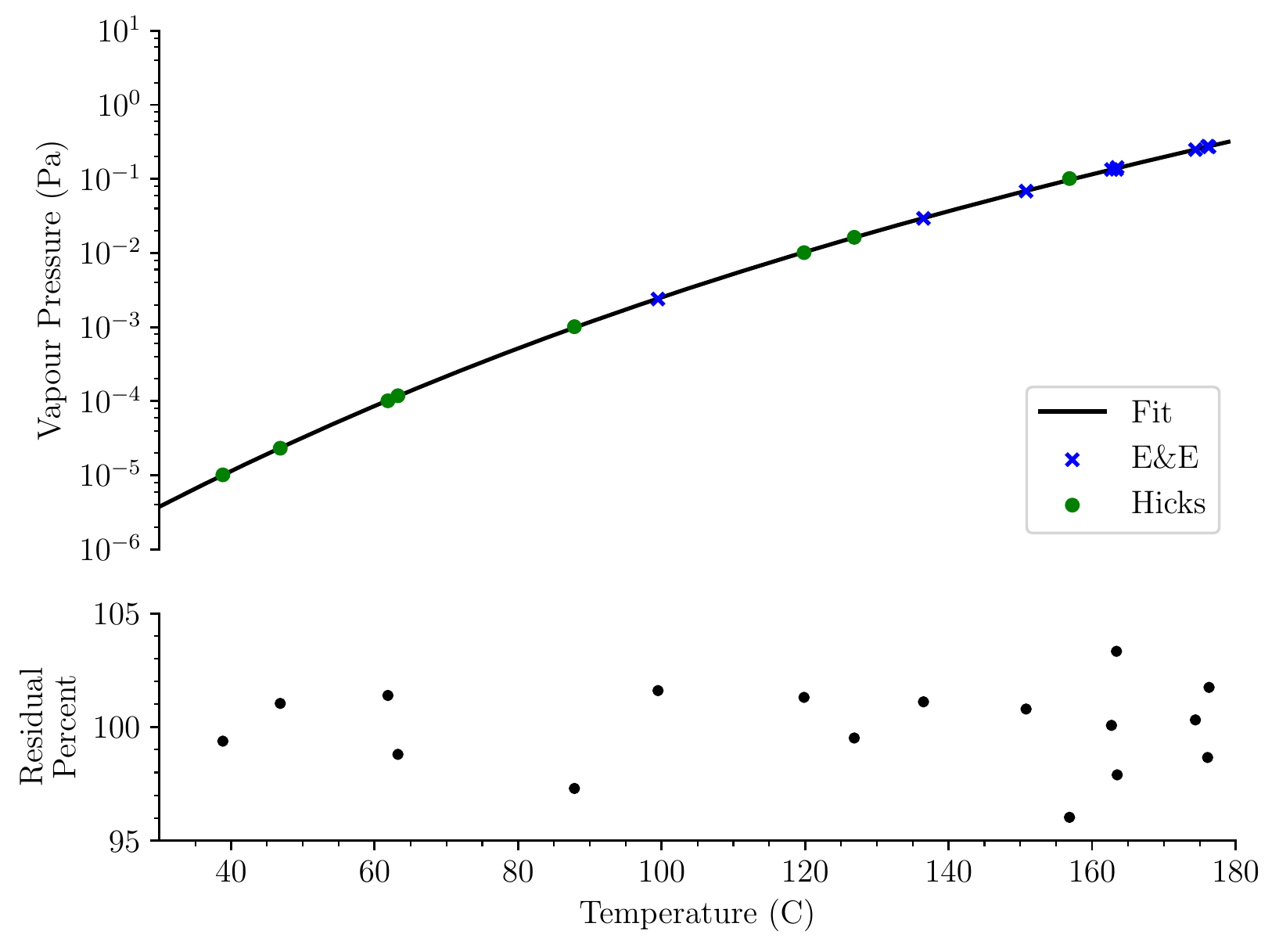}
  \caption[Potassium vapour pressure model]
          {Upper: The blue crosses and green dots show measurements of
          the vapour pressure of potassium made independently
          by~\citet{520.full} and~\citet{1.1733889} at various
          temperatures.  The black line shows a cubic fit to the data.
          Lower: The residuals to the fit in terms of percentage
          (i.e., $100 \cdot (\textrm{model}/\textrm{data})$).  The
          uncertainty on the source data are not available, and so no
          error bars are plotted.}
  \label{fig:vapour_model}
\end{figure*}

We already know the lifetime $\tau$ to be~\SI{27}{\nano\second}, the
atomic mass to be~\SI{39.0983}{\atomicmassunit}, the central
wavelength to be~\SI{769.898}{\nano\meter}, and it may be
shown~\citep[p.~97]{mitchell1961resonance} that the ratio $g_2/g_1$ is
unity.  The temperature of the reference cell is controlled, and so
the only remaining term is the number density.  We can determine the
number density based on the vapour pressure, since as we saw earlier
it depends only on the temperature and not on the amount of physical
material.

The vapour in a gas reference cell can be treated as an ideal gas
since the pressure is by design very low.  The cells are evacuated and
baked out before being filled with a small amount of potassium.  Since
the pressure is low, there will be few interactions between particles.
The ideal gas law is defined as,
\begin{equation}\label{eq:idealgas1}
  pV = N k_\mathrm{B} T ~,
\end{equation}
where $p$ is the pressure, $V$ the volume, $N$ the number of atoms or
molecules, $k_\mathrm{B}$ the Boltzmann constant, and $T$ the
temperature.  If we combine the number of atoms and the volume into
one parameter to form the number density $n(T)$, and define the
pressure in terms of temperature then equation~\ref{eq:idealgas1}
simplifies to,
\begin{equation}
  n(T) = \frac{p(T)}{k_\mathrm{B} T} ~,
\end{equation}
which depends only on the vapour pressure and the temperature.

Previous independent measurements have been
made~\citep{520.full,1.1733889} of the vapour pressure of potassium at
various temperatures.  These data have been used to fit a polynomial
in log space,
\begin{equation}\label{eq:vapour_model}
  p_\mathrm{vapour} = e^{a\,T^{3} + b\,T^{2} + c\,T + d} ~,
\end{equation}
where $p_\mathrm{vapour}$ is the vapour pressure and $T$ is the
temperature in Kelvin.  The values of the four fit coefficients are
shown in Table~\ref{table:vapour_model}.  A plot of results from this
model against the original data are shown in
Figure~\ref{fig:vapour_model}.  The model residuals are all
within~\SI{\pm5}{\percent} of the original value, with an upper limit
to the fit of~\SI{449}{\kelvin} (\SI{175}{\degreeCelsius}).  For a
typical vapour temperature of~\SI{100}{\degreeCelsius} the number
density required is~\SI{4.9e17}{\per\cubic\metre}.  This is equivalent
to~\SI{3.2e-8}{\kg} of potassium -- if you can see solid potassium in
the cell then there is enough to form the required vapour.

\begin{table}
    \centering
    \caption[Potassium vapour pressure model coefficients]
            {Potassium vapour pressure model coefficients.}
    \begin{tabular}{c S[table-format=-1.3e-1,round-mode=places,round-precision=3] S[table-format=-1.3e-1,round-mode=places,round-precision=3]}

    \toprule

    Coefficient & {Value} & {Uncertainty}\\

    \midrule

    a &  6.0054211235e-07 & 8.7468694167e-08\\
    b & -8.9121318942e-04 & 1.0072983768e-04\\
    c &  4.8897499745e-01 & 3.8399171195e-02\\
    d & -9.5551204080e+01 & 4.8436258495e+00\\

    \bottomrule

    \end{tabular}
    \label{table:vapour_model}
\end{table}

We now have everything in place to be able to calculate the optical
depth of a potassium vapour from equation~\ref{eq:od}.  We know all of
the parameters required to calculate the absorption coefficient
$\kappa_0$ in equation~\ref{eq:k0} by determining the vapour pressure
from equation~\ref{eq:vapour_model}, and we can calculate the
wavelength-dependent absorption coefficient from
equation~\ref{eq:dopplerabs}.  The model was verified by probing a
potassium vapour cell using a tunable diode laser.

\section{Laser Calibration}
\label{s:calibration}

\begin{figure*}[t]
  \centering
  \includegraphics[scale=1.0]{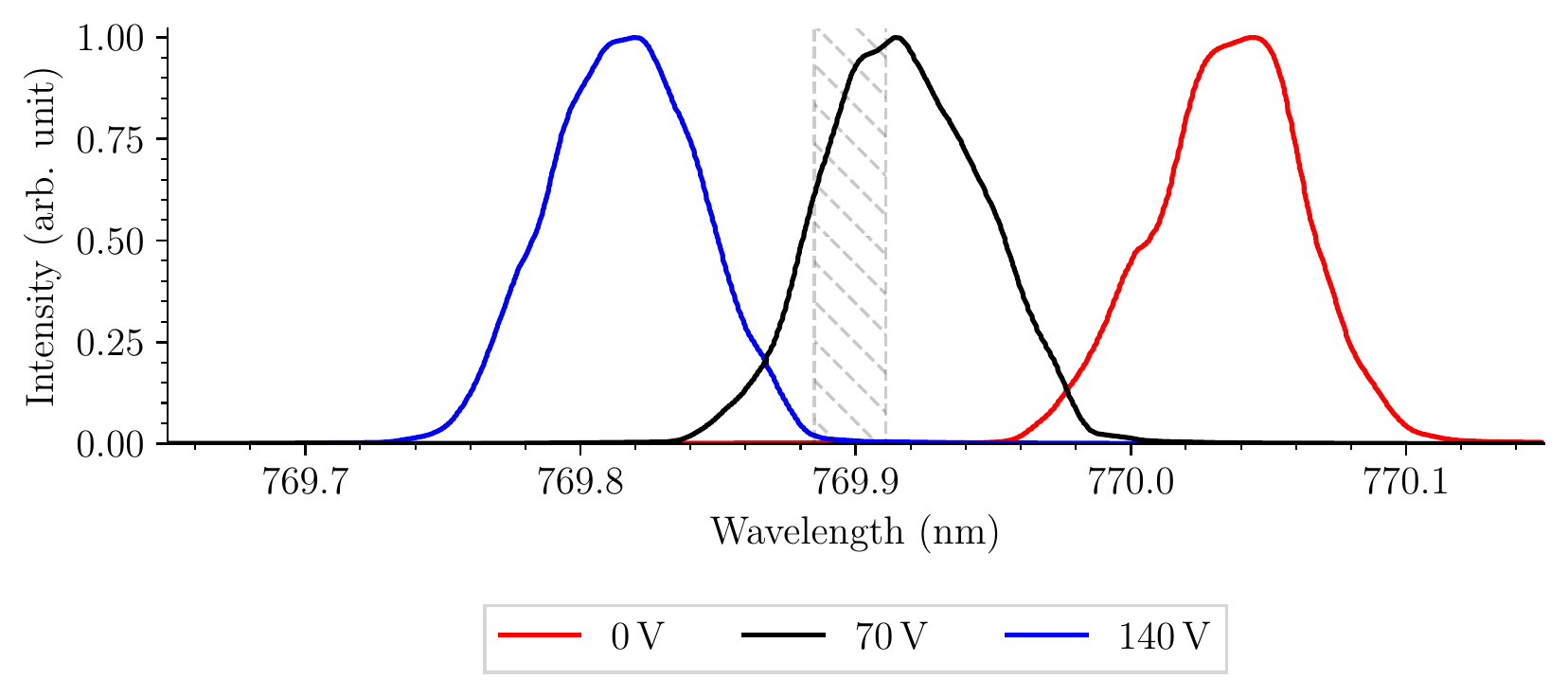}
  \caption[Toptica DLC\,pro tunable diode laser profile]
  {Toptica DLC\,pro tunable diode laser profile tuning range, measured
  using an Agilent 86142B optical spectrum analyser.  The line width is
  much narrower than indicated in the figure.  The measured FWHM is
  broadened to the minimum~\SI{0.06}{\nano\metre} resolution bandwidth
  of the spectrum analyser.  The red and blue lines indicate the
  extent of the voltage-controlled piezo fine-tuning range after prior
  coarse tuning to approximately~\SI{769.9}{\nano\metre}.  The hatched
  area indicates the extent of our wavelength region of interest.}
\label{fig:laser_range}
\end{figure*}

We used a Toptica DLC\,pro tunable diode laser, which is a Littrow-style
external cavity diode laser (EDCL)~\citep{toptica}.  The installed
laser diode has an emission spectrum and tuning range
of~\SIrange{765}{805}{\nano\meter}. The primary tuning mechanism is by
altering the angle of incidence of the grating forming the external
cavity, which for coarse adjustment is done using a micrometer screw,
and for fine adjustment by varying the applied voltage to a piezo
actuator.  Tuning can also be achieved by changing the diode drive
current and temperature.  The diode suffers mode-hopping as the tuning
parameters are adjusted, but this can be mitigated by varying all
parameters simultaneously.  The typical mode-hop-free tuning-range
(MHFTR) is quoted at~\SIrange{20}{50}{\giga\hertz}, which at a
wavelength of~\SI{770}{\nano\metre} is equivalent
to~\SIrange{40}{98}{\pico\metre}.  Achieving this range can be tricky
due to the interaction of all the various tuning methods.  The typical
line width achieved by a Littrow-style ECDL over a duration
of~\SI{5}{\micro\second} is~\SIrange{10}{300}{\kilo\hertz}, which
again at a wavelength of~\SI{770}{\nano\metre} is less
than~\SI{0.6}{\femto\metre}.  The main contributions to the line width
are electronic noise, acoustic noise, and other vibrations that affect
the cavity length.

\begin{figure*}[t]
  \centering
  \includegraphics[scale=1.0]{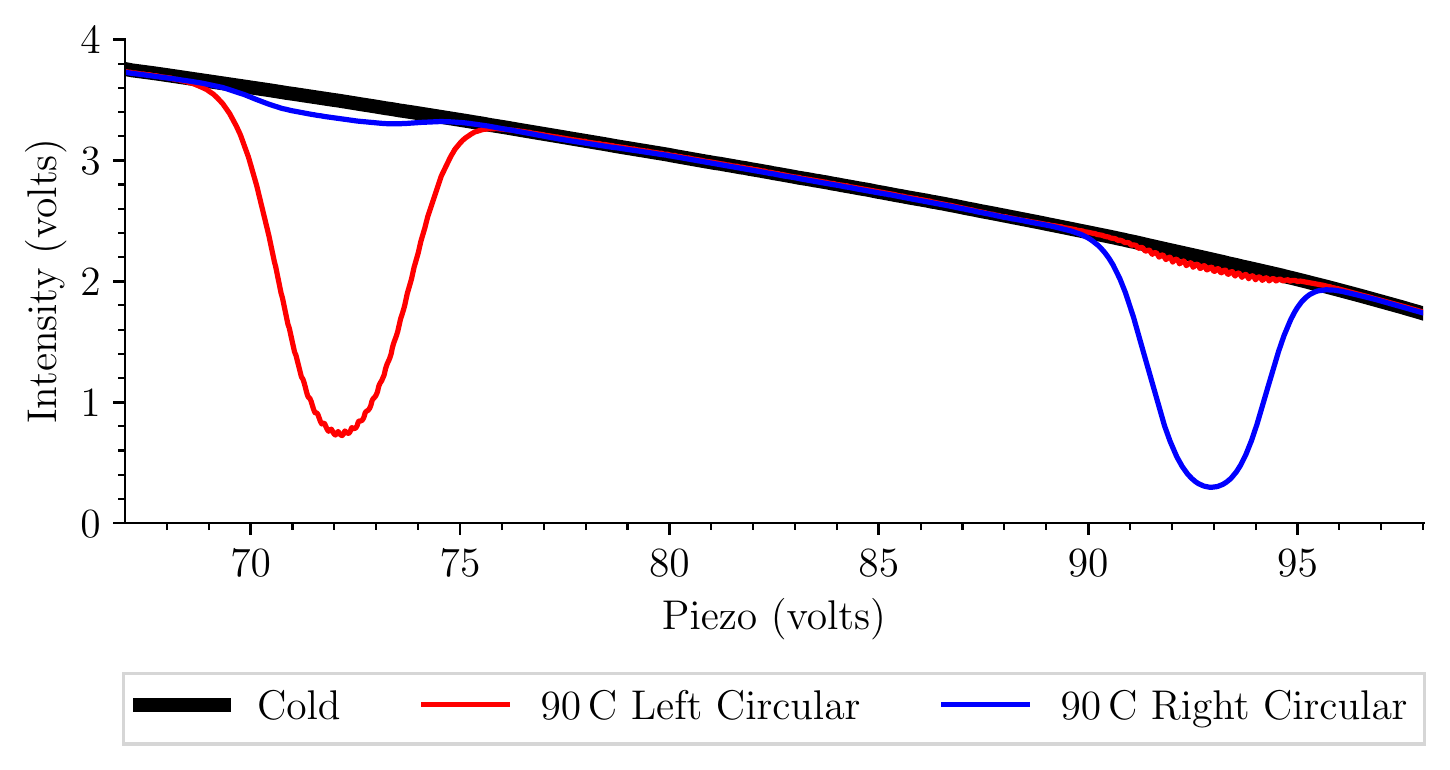}
  \caption[Laser intensity flat-field calibration]
  {Laser intensity flat-field calibration.}
  \label{fig:calibration1}
\end{figure*}

The output of the ECDL has no intrinsic wavelength calibration, except
that the wavelength must be somewhere within the broadband output of
the model of installed laser diode.  In order to coarse-tune the laser
to approximately~\SI{769.9}{\nano\metre} the output wavelength was
measured using an Agilent 86142B optical spectrum
analyser~\citep{agilent_technical}, shown in
Figure~\ref{fig:laser_range}.  The 86142B has a resolution bandwidth
(RBW) of~\SI{0.06}{\nano\metre} and so the measured laser line width
is considerably broadened.  This resolution is sufficient to allow use
of the coarse manual micrometer screw adjustment to ensure the
required wavelength is positioned within the narrow scan range of the
piezo actuator.  The resolution is insufficient to tune the wavelength
precisely to the desired absorption line, but once coarse calibration
is achieved simply scanning between the piezo limits allows the
absorption line to be found.  In order to calibrate the width and
depth of the absorption line at varying temperature, it is necessary
to calibrate the piezo voltage in terms of wavelength and to normalise
the beam intensity.  This calibration is non-trivial and we will now
go on to discuss the steps required to achieve the final overall
calibration.

\begin{figure*}[t]
  \centering
  \includegraphics[scale=1.0]{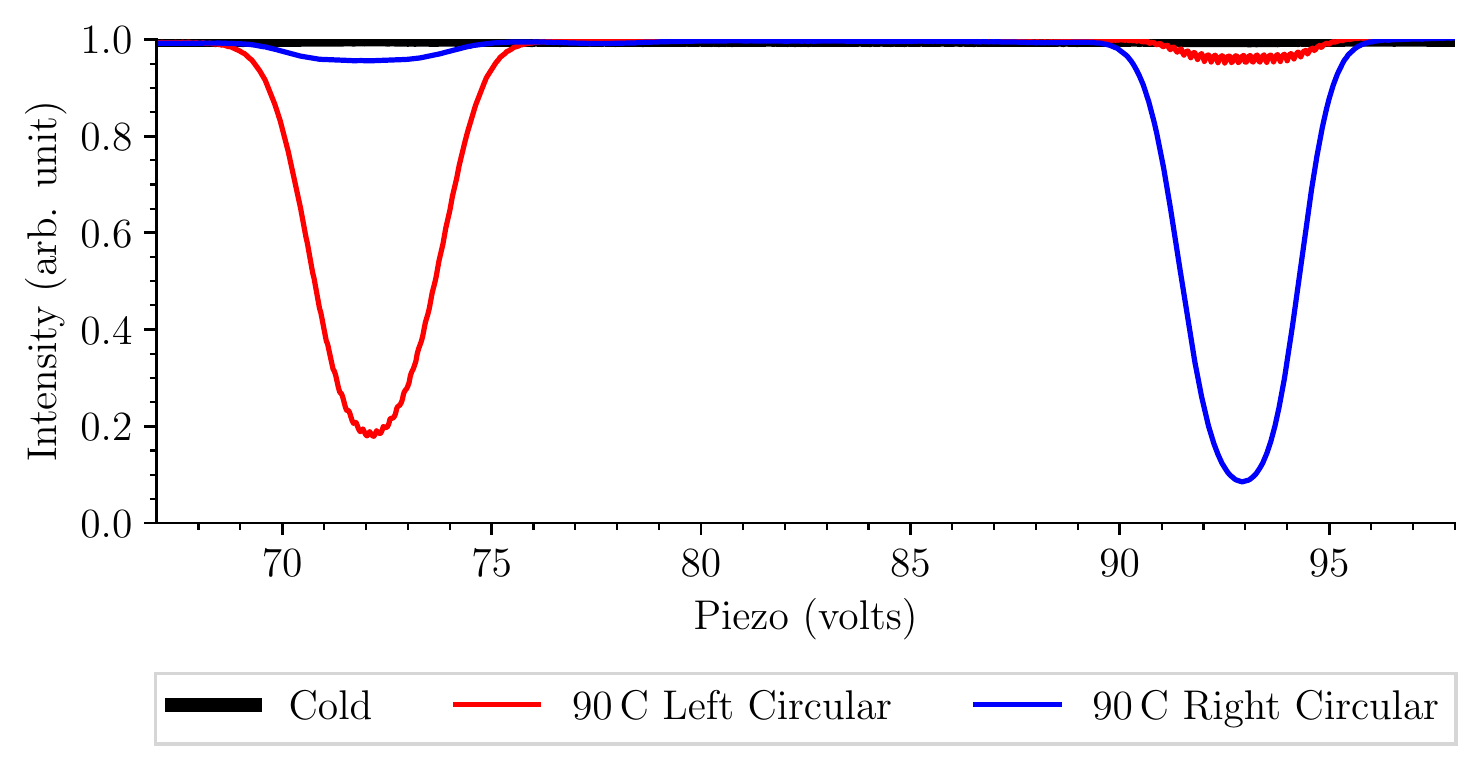}
  \caption{Piezo voltage calibration after flat-field adjustment.  The
    line centres occur at~\SI{72.2}{\volt} and \SI{92.9}{\volt}.  The
    different depths are due to slight misalignment of the polarising
    filters.  There is also some evidence of non-linearity in the
    wavelength scan observed from slight differences in the profile
    widths.}
  \label{fig:calibration2}
\end{figure*}

Since BiSON spectrophotometers use a magnetic field to Zeeman split
the reference absorption line and produce two separate instrumental
passbands, this technique can be used to produce a known wavelength
separation and so calibrate the piezo.  Figure~\ref{fig:calibration1}
shows the two components of the Zeeman split absorption line with the
cell heated to~\SI{90}{\celsius} in a~\SI{0.3}{\tesla} longitudinal
magnetic field.  The beam intensity is not uniform across the piezo
scan due to a drive-current feed-forward mechanism employed to
increase the MHFTR of the diode laser.  It is imperative that the
diode does not mode-hop during the scan since this would result in
different wavelength calibrations either side of the mode-hop,
destroying the overall calibration.  The feed-forward adjusts the
drive-current proportionally with the piezo position-voltage by a
factor of~\SI{-0.32}{\milli\ampere\per\volt}, which helps prevent
mode-hopping but has the undesired side-effect of changing the beam
intensity.  If a mode-hop continues to appear at an inconvenient
wavelength, then the temperature of the diode can be adjusted to move
the mode-hop elsewhere.  The mean drive-current and detector-gain has
to be selected carefully to ensure that the current always stays above
the lasing threshold of~\SI{109}{\milli\ampere}, but also that the
beam intensity does not exceed the dynamic range of the detector
throughout the whole piezo scan range.  Once a mode-hop-free scan has
been achieved with a beam intensity that is comfortably within the
dynamic range of the detector, a flat-field can be captured and used
to normalise the intensity of the scan, shown in
Figure~\ref{fig:calibration2}.

\begin{figure*}[t]
  \centering
  \includegraphics[scale=1.0]{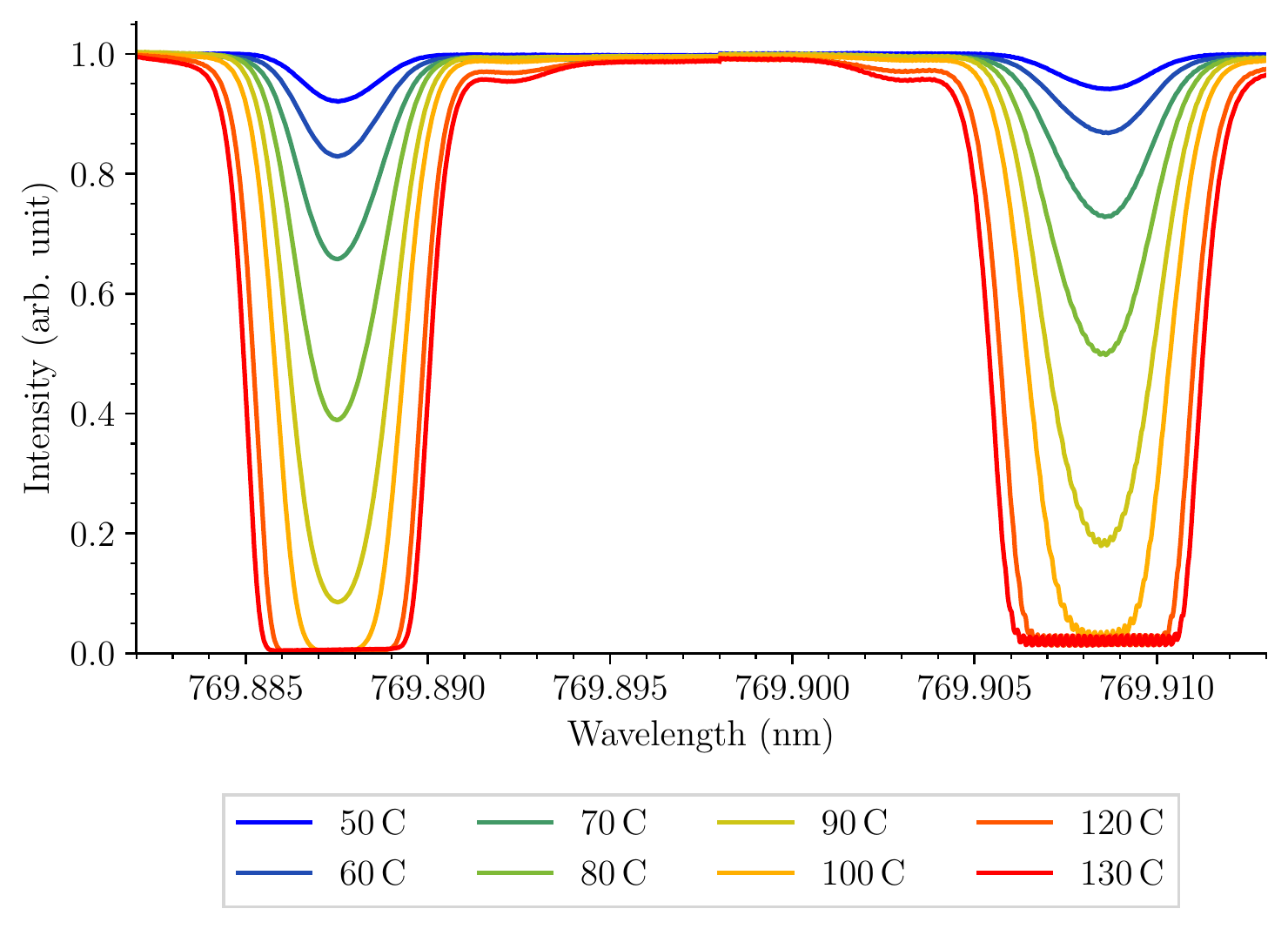}
  \caption[Vapour cell absorption intensity profiles]
  {Vapour cell absorption intensity profiles at a range of
  temperatures.}
  \label{fig:transmission}
\end{figure*}

We can complete the calibration by making use of the known position of
the two Zeeman components, and applying the splitting to the piezo
voltages at the measured line centres.  The calibrated absorption
profiles for eight different cell temperatures are shown in
Figure~\ref{fig:transmission}.  There is some evidence of slight
misalignment of the polarising filters indicated by differences in
absorption depth between the two polarisation states.  There is also
some non-linearity in the wavelength scan indicated by a slight
difference in width between the two polarisation states.  These errors
are small and will be mitigated against by fitting the model to both
absorption components simultaneously, allowing all equal-and-opposite
errors to cancel.

\section{Model Validation}
\label{s:fitting}

\begin{figure*}[t]
  \centering
  \includegraphics[scale=1.0]{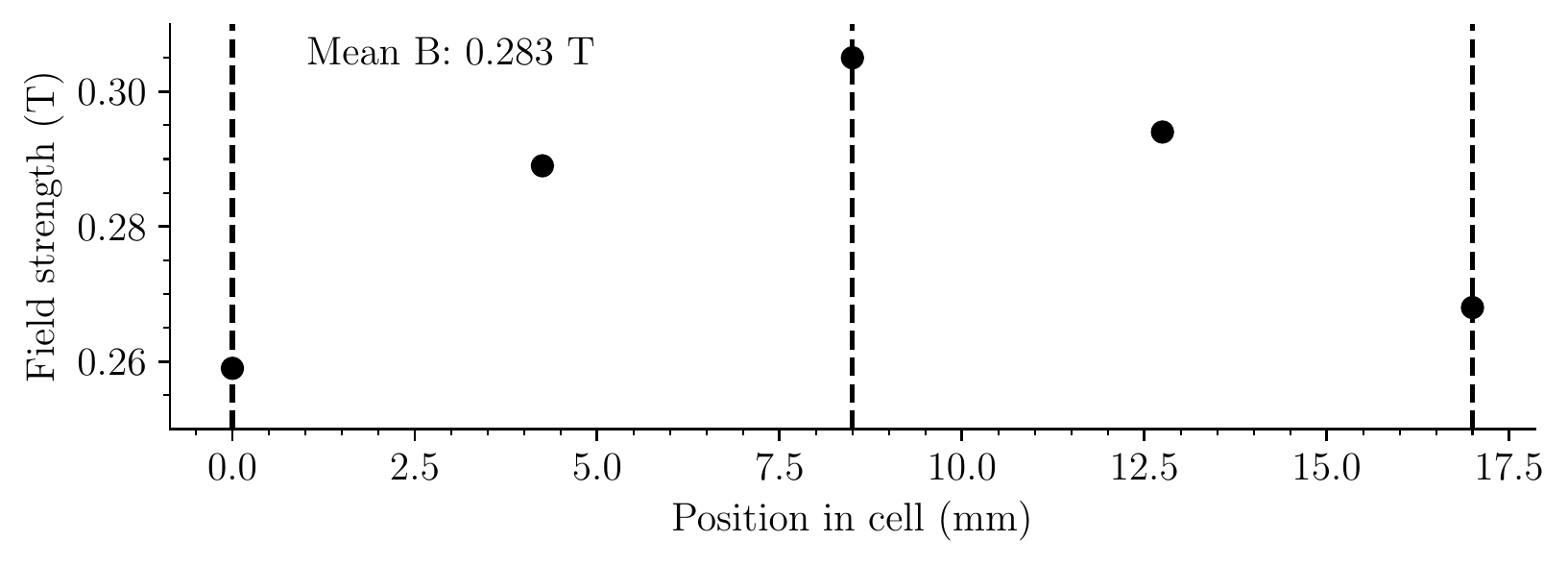}
  \caption{On-axis magnetic variation inside the cell oven.  The field
    inhomogeneity adds an additional source of spectral broadening
    which must be included in the model.}
  \label{fig:magnet}
\end{figure*}

\begin{figure*}[p]
  \centering
  \includegraphics[scale=1.0]{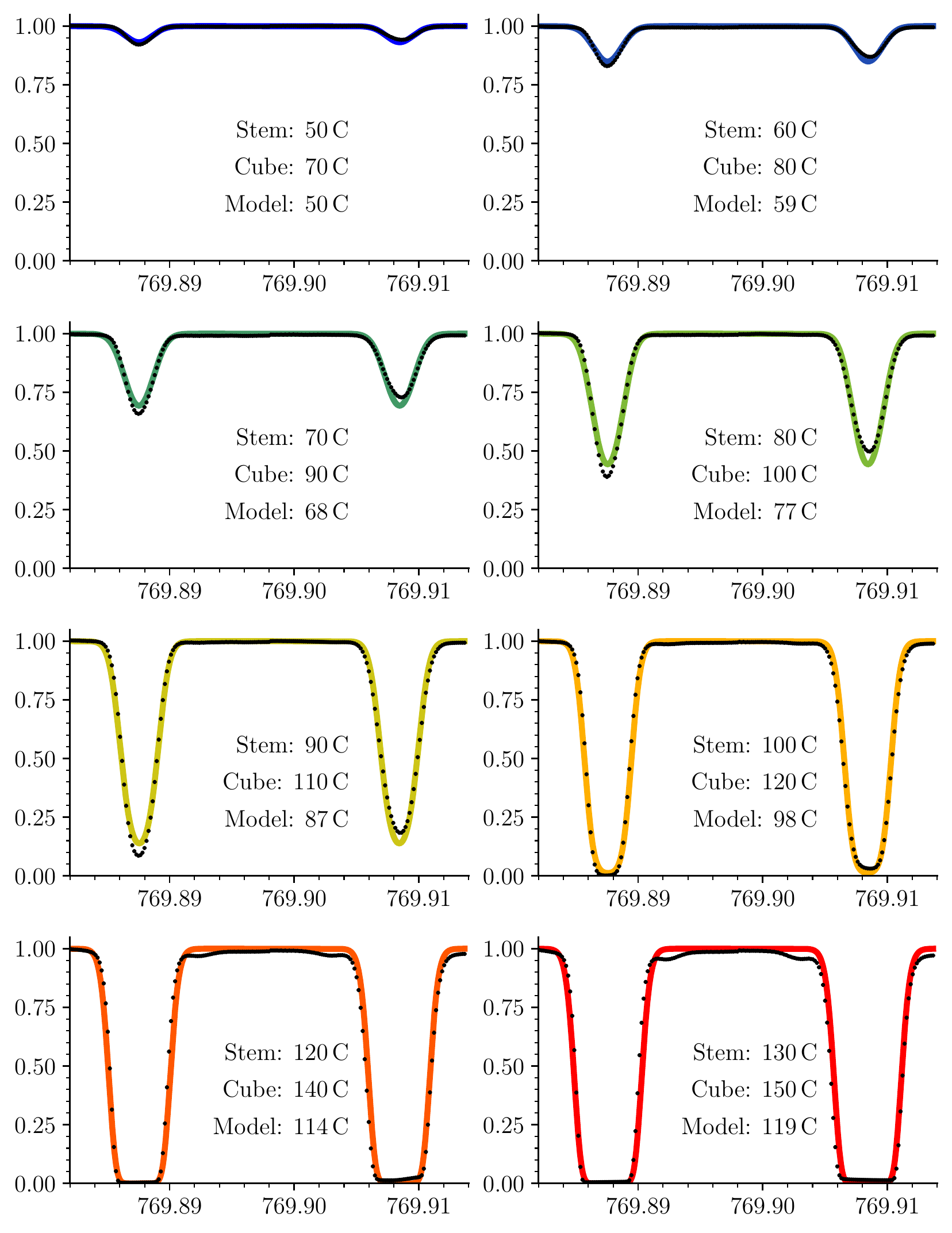}
  \caption[Modelled absorption profiles at a range of vapour
    temperatures] {Modelled absorption profiles at a range of vapour
    temperatures.  The dotted lines show the measured absorption
    profile.  The solid lines show the results of the fit to the data.
    Both absorption components are fitted simultaneously allowing
    systematic equal-and-opposite errors to cancel.}
  \label{fig:model}
\end{figure*}

\begin{table*}
    \centering
    \sisetup{table-format=3.1,round-minimum=0.1,round-mode=places,round-precision=1}
    \caption{Fitted model parameters at a range of vapour temperatures.}
    \begin{tabular}{S[table-format=3.0]
                    S[table-format=3.0]
                    S[table-format=3.0,round-precision=0]
                    S[table-format=<1.1]
                    S[table-format=-2.1]
                    S[table-format=1.2,round-precision=2]
                    S[table-format=<1.2,round-precision=2,round-minimum=0.01]}

    \toprule
        \multicolumn{2}{c}{{\bfseries Oven Temp}} & \multicolumn{3}{c}{{\bfseries Model Temp}} & \multicolumn{2}{c}{{\bfseries Broadening Factor}}\\
        \cmidrule(lr){1-2}\cmidrule(lr){3-5}\cmidrule(lr){6-7}
        {Stem} & {Cube} & {Stem} & {Uncertainty} & {$\Delta T$} & {Scale} & {Uncertainty}\\
    \midrule

     50 &  70 &  49.93801683529483 & 0.05911385274160247  &  -0.06198316470516829 & 1.5497977516900034 & 0.010225235529756882\\
     60 &  80 &  59.07193440488454 & 0.05806678516341524  &  -0.9280655951154628  & 1.593463750618715  & 0.009660575562317709\\
     70 &  90 &  67.8763995136866  & 0.05387503498457559  &  -2.123600486313407   & 1.551258089828436  & 0.008040919986667729\\
     80 & 100 &  76.78513535292292 & 0.049740470851325234 &  -3.2148646470770785  & 1.4761742475672481 & 0.006132750128753006\\
     90 & 110 &  87.3266062018605  & 0.06018561298799116  &  -2.6733937981395     & 1.3954584936066434 & 0.004645848115386062\\
    100 & 120 &  97.80252696851991 & 0.14375630603606684  &  -2.1974730314800865  & 1.3611303163197057 & 0.005682520990971153\\
    120 & 140 & 114.12485974429399 & 0.605734316440814    &  -5.875140255706015   & 1.4125803194248847 & 0.01311016578462419\\
    130 & 150 & 119.23633188810875 & 0.8460268659074462   & -10.76366811189125    & 1.4694733287114523 & 0.01670930894195373\\
    
    \bottomrule

    \end{tabular}
    \label{table:fitted_temperatures}
\end{table*}

The laser calibration is dependent on the accuracy and uniformity of
the vapour cell longitudinal magnetic field.  The field strength at
several points inside the cell oven was measured using a Hirst GM04
Gaussmeter, which is factory calibrated using nuclear magnetic
resonance to correct for irregularities in the supplied semi-flexible
transverse Hall probe.  It was found that the target field strength
of~\SI{0.3}{\tesla} is achieved only at the cell centre and drops to
approximately~\SI{0.26}{\tesla} at the front and rear of the cell
oven, as shown in Figure~\ref{fig:magnet}.  The mean field strength
is~\SI{0.283}{\tesla}.  Due to the variation in field strength, it is
not necessary to measure the field strength to high precision, however
this does mean that the absorption profile will have an additional
broadening factor caused by the non-uniform magnetic field, and the
FWHM determined from equation~\ref{eq:doppler6} has to be modified to
allow an additional degree of freedom.

The resulting final model has been fitted against absorption profiles
for all eight vapour temperatures in Figure~\ref{fig:transmission}.
Both absorption components are fitted simultaneously allowing
systematic equal-and-opposite errors to cancel.  The results are shown
in Figure~\ref{fig:model}, where the dotted lines show the measured
absorption profile and the solid lines show the results of the fit to
the data.  The fitted parameters and uncertainties listed in
Table~\ref{table:fitted_temperatures}.

All fitted temperatures indicate that the vapour pressure is
controlled by the temperature of the solid potassium in the stem of
the cell.  The temperature of the upper cube of the cell simply has to
be higher than the stem and high enough to ensure that deposition of
solid potassium onto the windows does not occur.  The increasing
difference between the model temperature and the stem temperature at
higher values is an indication of the difficulties of coupling heat
efficiently into the cell.  Heating of the stem is achieved by a foil
heating element wrapped around the glass, and heat does of course leak
into the surrounding chassis as well as heating the cell.  At higher
temperatures the losses become more significant and it becomes
progressively more difficult to effectively heat the cell, and this is
indicated by the real temperature of the vapour as determined from the
model being a few degrees below the measured temperature of the stem
heating element.

\section{Instrumentation Parameters}

\label{s:detector}
\begin{figure}[t]
  \centering
  \includegraphics[scale=1.0]{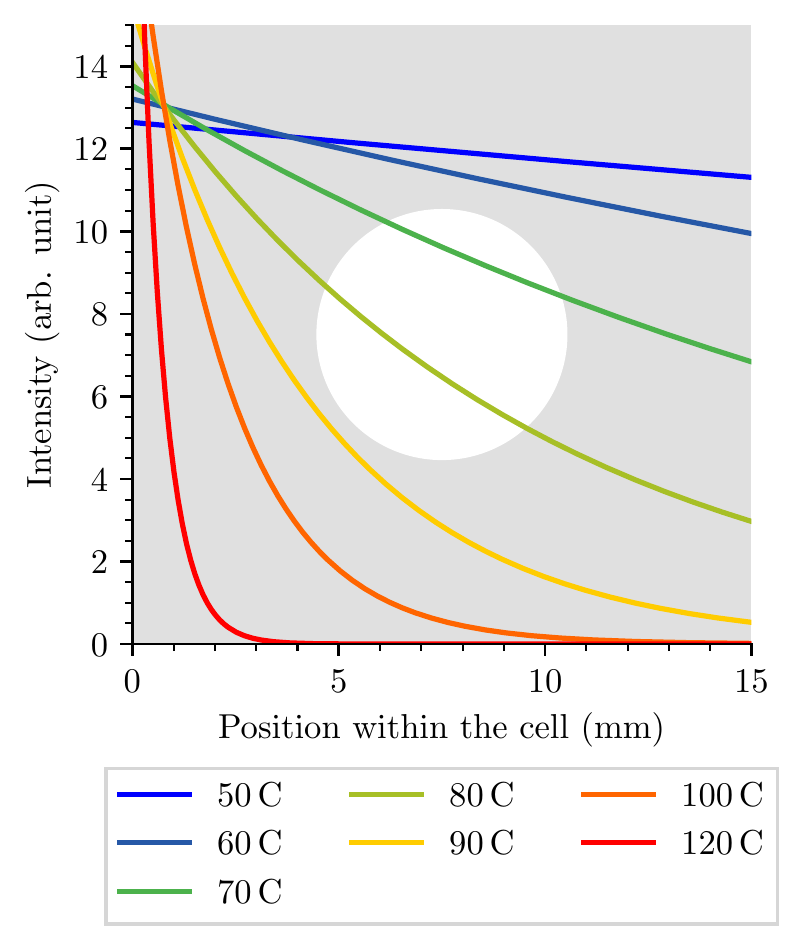}
  \caption[Absorption through a~\SI{15}{\milli\metre} vapour cell at various temperatures]
  {Absorption through a typical~\SI{15}{\milli\metre} BiSON vapour
  cell at various temperatures.  The absorption coefficient is
  modelled as monochromatic but with the initial intensity scaled by
  the expected FWHM to compensate for the differences in spectral
  bandwidth.  The white circle in the centre indicates the
  typical~\SI{6}{\milli\metre} aperture of a scattering detector.}
  \label{fig:bison_cell}
\end{figure}

\begin{figure*}
  \centering
  \includegraphics[scale=1.0]{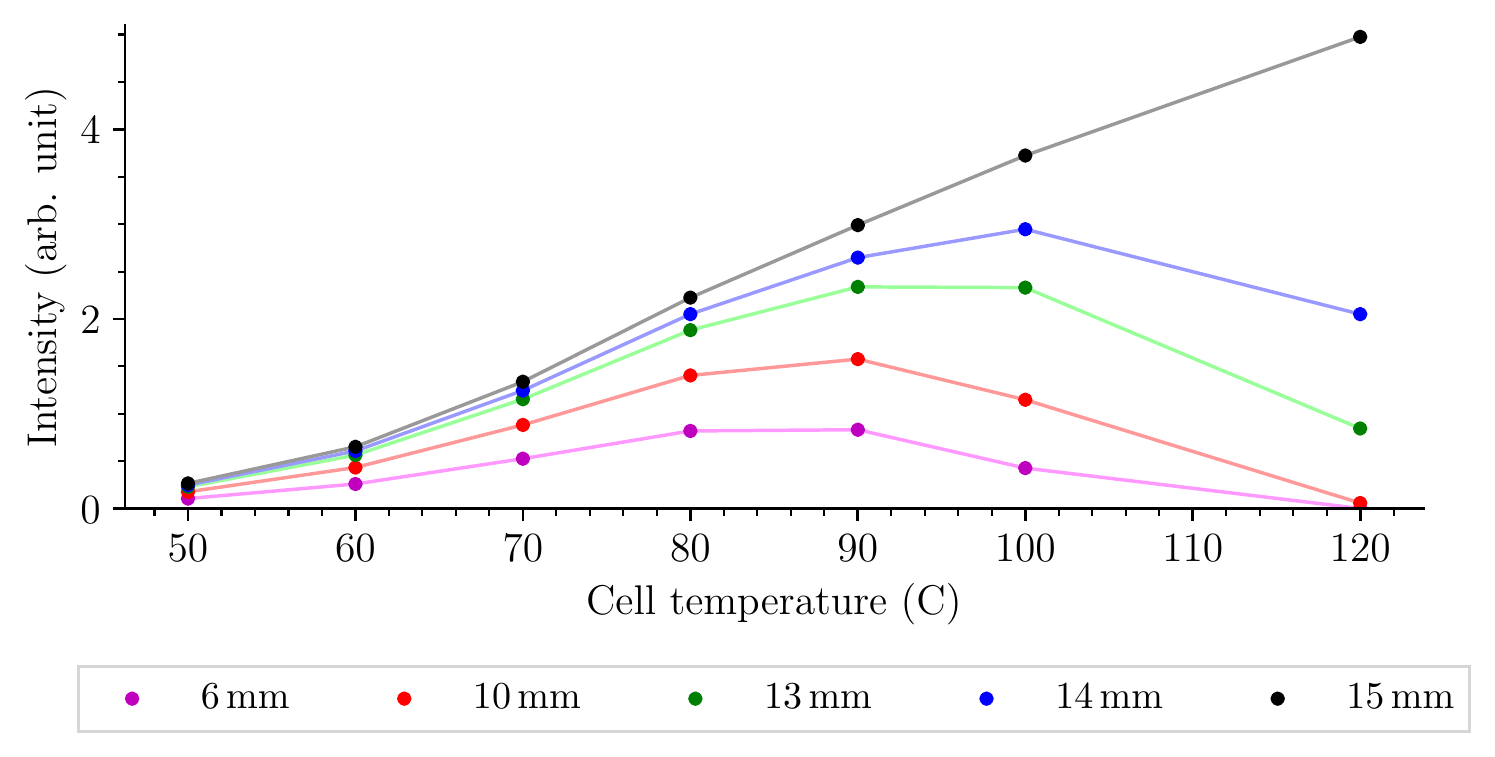}
  \caption[Optimum vapour temperature at various apertures]
  {Optimum vapour temperature for maximum scattering intensity at a
  range of aperture sizes.}
  \label{fig:scatter_optimise}
\end{figure*}

The model can be used to estimate the scattering performance of the
cell and the effect of different detector aperture sizes and
positions.  Figure~\ref{fig:bison_cell} shows the expected intensity
change as light passes through the cell determined from
equation~\ref{eq:absorptioncoeff}.  The absorption coefficient is
modelled as monochromatic from equation~\ref{eq:k0} but with the
initial intensity scaled by the expected FWHM to compensate for the
differences in spectral bandwidth.  We make an assumption that any
photon is scattered only once.  Any secondary (or further multiple)
scattering of light that has been moved away from the optical axis in
the direction of a scattering detector will be isotropic. The detected
intensity will be reduced by the optical depth between the point of
scattering within the cell and the detector, and so it is important
when calculating the total expected throughput, but has only a minor
effect when considering variations in aperture size.  The white circle
in the centre of the figure indicates the typical~\SI{6}{\milli\metre}
aperture of a scattering detector, with scattered light perpendicular
to the page coming out towards the reader as the beam passes through
from left to right. It is clear that at low temperatures there is very
little scattered light due to the small absorption coefficient.  At
high temperatures almost all the light is scattered at the front of
the cell and so not captured by the detector.  This implies that there
will be some optimum temperature at which the measured scattering
intensity is greatest.  By considering the change in intensity within
the area of the white circle, we can estimate the intensity of light
scattered away from the beam optical axis.
Figure~\ref{fig:scatter_optimise} shows the estimated scattering
intensity for several aperture sizes and vapour temperatures.  If the
aperture is equal to the size of the vapour cell then all the light
scattered towards the detector is captured, and the intensity
continually increases with temperature.  As the aperture size is
reduced a temperature plateau forms.  One might expect that the ideal
aperture is as large as possible in order to maximise the intensity of
scattered light and so maximise the signal-to-noise ratio.  However,
there are several reasons why this is not the case.  Applying a
smaller aperture and causing the temperature plateau to form minimises
the effect of fluctuations in vapour temperature, and the
temperature-width of the plateau is greatest at smaller apertures.
Also, it is imperative to eliminate non-resonantly scattered light
from reaching the detector, such as specular reflections from the cell
glass especially at the edges and corners of the cell, and again this
is best achieved via small apertures effectively acting as a baffle.

\begin{figure*}
  \centering
  \includegraphics[scale=1.0]{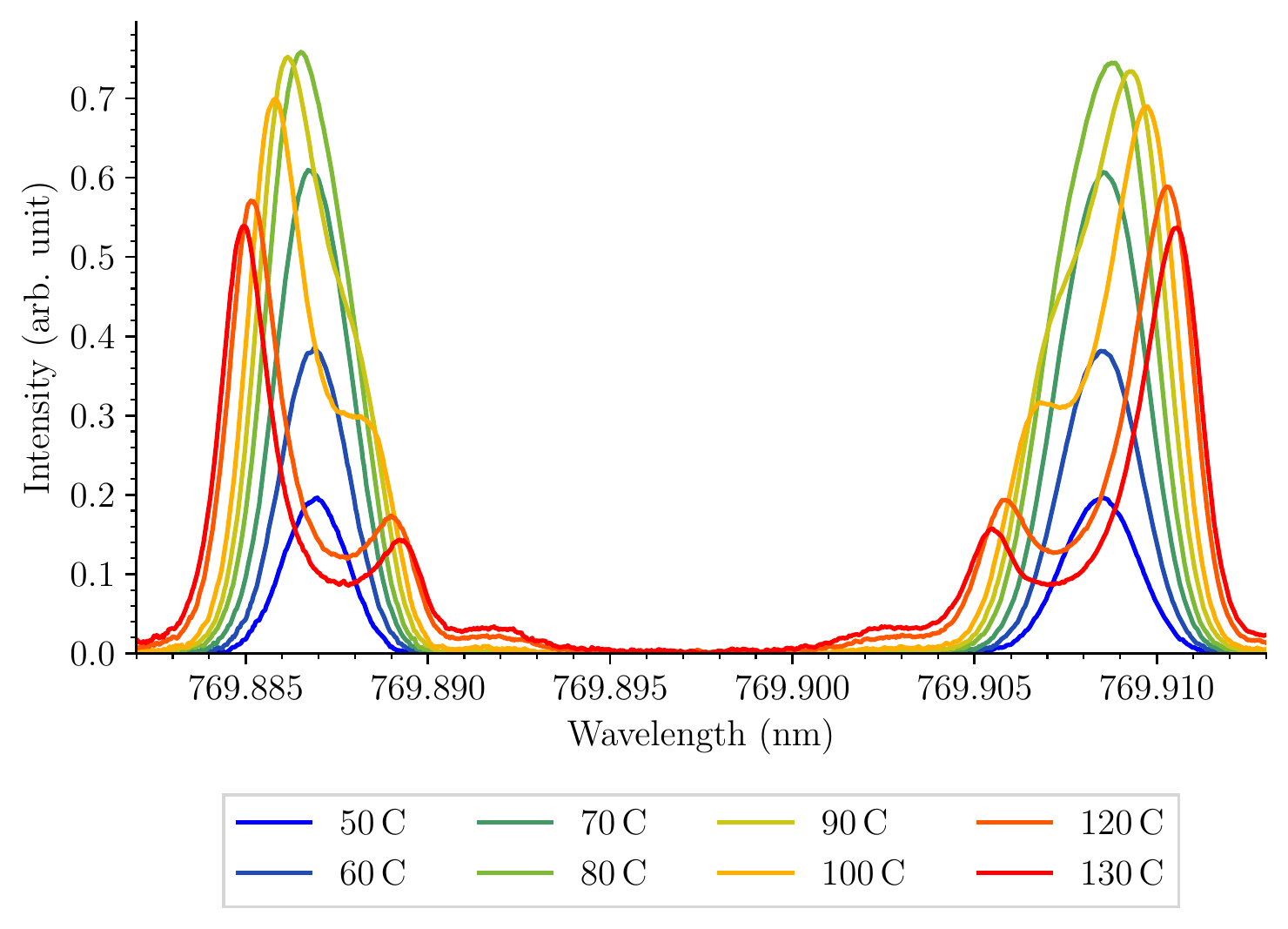}
  \caption[Vapour cell scattering intensity profiles]
  {Vapour cell scattering intensity profiles at a range of
  temperatures. The profile asymmetry seen at temperatures above the
  ideal temperature is a result of the inhomogeneity of the magnetic
  field.}
  \label{fig:scattering}
\end{figure*}

The scattering detectors used in most BiSON instrumentation
employ~\SIrange{6}{10}{\milli\metre} circular apertures which suggests
that the maximum intensity will be found at a stem temperature of
approximately~\SIrange{80}{90}{\celsius}.  Many {BiSON} instruments
run at a stem temperature of~\SI{90}{\celsius} and a cube temperature
of~\SI{110}{\celsius} and so these results agree with expectation
based on experience.  At this temperature the optical depth is
approximately one if measured from the front of the cell to the
centre, or approximately two from the front of the cell to the rear.

Figure~\ref{fig:scattering} shows the scattering profiles taken
simultaneously with the absorption profiles shown in
Figure~\ref{fig:transmission} over a range of vapour cell
temperatures, and the intensity peaks at
approximately~\SIrange{80}{90}{\celsius} as expected.  The profile
asymmetry seen at high temperatures is a result of the inhomogeneity
of the magnetic field. At lower temperatures, light is detected from
across the whole aperture, and the peaks are centred at the
wavelengths corresponding to the splitting produced by the mean field
strength.  At higher temperatures, the scattering is dominated towards
the front of the cell within the weaker part of the field, and not
seen by the detector.  The detector captures only the remaining small
fraction of light at the wavelength corresponding to the higher field
strength at the centre of the cell, leading to the observed offset
weighted towards higher field strength.

\section{Conclusion}

We have developed a model of a {BiSON} potassium vapour cell and
validated the output of the model using a tunable diode laser, where
the model was found to effectively reproduce the properties of the
cell.  Previously the ideal temperature working point was determined
empirically, and the result found here now puts the choice on a firm
footing.

The model can be used to assess vapour cells of different size and
shape, and determine the ideal operating vapour temperature and
configuration of detectors, allowing comparison of expected
performance between existing {BiSON} vapour cells and commercially
available alternatives.  Previously the optical depth of the cell, and
the location within the cell where most of the scattering occurs, has
not been known precisely.  This information is essential when
designing new instrumentation, such as the prototype next generation
of BiSON spectrophotometers ({BiSON:NG}\,\citep{halephd}), where the
aim is to make use of off-the-shelf components to simplify and
miniaturise the instrumentation as much as practical.

\section{Open Data}

All code and data are freely available for download from the
University of Birmingham eData archive~\citep{edata417}, and also from
the source GitLab repository~\citep{gitlab}.

\ack
The authors are grateful for the financial support of the Science and
Technology Facilities Council (STFC), grant reference ST/R000417/1,
and to the College of Engineering and Physical Sciences at the
University of Birmingham for the purchase of the DLC\,pro tunable diode
laser.

\bibliographystyle{unsrtnat}
\bibliography{references}

\end{document}